\documentclass[10pt,A4paper,conference]{IEEEtran}
\usepackage{amsthm}
\usepackage{amsmath}
\usepackage{amsfonts}
\usepackage{graphicx}
\usepackage{latexsym}
\usepackage{amssymb}
\usepackage{stmaryrd}
\usepackage{amscd}
\usepackage[small]{caption}

\newcommand{\deff}{\mbox{$\stackrel{\rm def}{=}$}}

\newcommand{\sbinom}[2]{\left[ \begin{array}{c} #1 \\ #2 \end{array} \right] }

\newcommand{\field}[1]{\mathbb{#1}}

\newcommand{\cA}{{\cal A}}

\newcommand{\cC}{{\cal C}}
\newcommand{\cD}{{\cal D}}

\newcommand{\cS}{{\cal S}}

\newcommand{\cV}{{\cal V}}

\newcommand{\sP}{\field{P}}

\newcommand{\sG}{\field{G}}

\DeclareMathAlphabet{\mathbfsl}{OT1}{cmr}{bx}{it}
\newcommand{\uuu}{\kern-1pt\mathbfsl{u}\kern-0.5pt}
\newcommand{\vvv}{\kern-1pt\mathbfsl{v}\kern-0.5pt}

\newcommand{\myboxplus}{\kern1pt\mbox{\small$\boxplus$}}

\makeatletter \DeclareRobustCommand{\sbinom}{\genfrac[]\z@{}}
\makeatother
\newcommand{\G}[2]{\sbinom{{#1}\kern-1pt}{{#2}\kern-1pt}}
\newcommand{\Gq}[2]{\sbinom{{#1}\kern-0.25pt}{{#2}\kern-0.25pt}}

\newcommand{\Ps}{\smash{{\sP\kern-2.0pt}_q\kern-0.5pt(n)}}
\newcommand{\sPs}{\smash{{\sP\kern-1.5pt}_q(n)}}
\newcommand{\Ptwo}{\smash{{\sP\kern-2.0pt}_2\kern-0.5pt(n)}}
\newcommand{\Ptwom}{\smash{{\sP\kern-2.0pt}_2\kern-0.5pt(m)}}
\newcommand{\Ptwonm}{\smash{{\sP\kern-2.0pt}_2\kern-0.5pt(n+m)}}
\newcommand{\Ptwoa}{\smash{{\sP\kern-2.0pt}_2\kern-0.5pt(1)}}
\newcommand{\Ptwob}{\smash{{\sP\kern-2.0pt}_2\kern-0.5pt(2)}}
\newcommand{\Ptwoc}{\smash{{\sP\kern-2.0pt}_2\kern-0.5pt(3)}}
\newcommand{\Ptwod}{\smash{{\sP\kern-2.0pt}_2\kern-0.5pt(4)}}
\newcommand{\Ptwoe}{\smash{{\sP\kern-2.0pt}_2\kern-0.5pt(5)}}
\newcommand{\Ptwof}{\smash{{\sP\kern-2.0pt}_2\kern-0.5pt(6)}}
\newcommand{\Ptwokm}{\smash{{\sP\kern-2.0pt}_2\kern-0.5pt(2k-1)}}
\newcommand{\Pone}{\smash{{\sP\kern-2.5pt}_2\kern-0.5pt(n{-}1)}}

\newcommand{\Gr}{\smash{{\sG\kern-1.5pt}_q\kern-0.5pt(n,k)}}
\newcommand{\Gi}{\smash{{\sG\kern-1.5pt}_q\kern-0.5pt(n,i)}}
\newcommand{\Gj}{\smash{{\sG\kern-1.5pt}_q\kern-0.5pt(n,j)}}
\newcommand{\Grmk}{\smash{{\sG\kern-1.5pt}_q\kern-0.5pt(n,n-k)}}
\newcommand{\Grdk}{\smash{{\sG\kern-1.5pt}_q\kern-0.5pt(2k,k)}}
\newcommand{\Grekappa}{\smash{{\sG\kern-1.5pt}_q\kern-0.5pt(n,e+1-\kappa)}}
\newcommand{\Grtwoekappa}{\smash{{\sG\kern-1.5pt}_q\kern-0.5pt(n,2e+1-\kappa)}}
\newcommand{\Gremkappa}{\smash{{\sG\kern-1.5pt}_q\kern-0.5pt(n,e-\kappa)}}
\newcommand{\Gn}{\smash{{\sG\kern-1.5pt}_2\kern-0.5pt(n,n{-}1)}}
\newcommand{\Gnq}{\smash{{\sG\kern-1.5pt}_q\kern-0.5pt(n,n{-}1)}}
\newcommand{\Gone}{\smash{{\sG\kern-1.5pt}_2\kern-0.5pt(n,1)}}
\newcommand{\Gqone}{\smash{{\sG\kern-1.5pt}_q\kern-0.5pt(n,1)}}
\newcommand{\GTwo}{\smash{{\sG\kern-1.5pt}_2\kern-0.5pt(n,k)}}
\newcommand{\GTwonk}[2]{{\smash{{\sG\kern-1.5pt}_2\kern-0.5pt({#1},{#2})}}}
\newcommand{\Gnk}{\smash{{\sG\kern-1.5pt}_2\kern-0.5pt(n,n{-}k)}}
\newcommand{\Greone}{\smash{{\sG\kern-1.5pt}_q\kern-0.5pt(n,e{+}1)}}
\newcommand{\Gretwo}{\smash{{\sG\kern-1.5pt}_q\kern-0.5pt(n,e{+}2)}}

\newcommand{\be}[1]{\begin{equation}\label{#1}}
\newcommand{\ee}{\end{equation}}

\newcommand{\Cref}[1]{Co\-rol\-la\-ry\,\ref{#1}}

\newcommand\scalemath[2]{\scalebox{#1}{\mbox{\ensuremath{\displaystyle #2}}}}

\newtheorem{theorem}{Theorem}
\newtheorem{lemma}{Lemma}

\newtheorem{cor}{Corollary}

\begin{document}

\title{Perfect Permutation Codes with \\ the Kendall's $\tau$-Metric\vspace{-1.0ex}}

\author{\authorblockN{Sarit Buzaglo}
\authorblockA{Dept. of Computer Science\\
Technion-Israel Institute of Technology\\
Haifa 32000, Israel \\
Email: sarahb@cs.technion.ac.il} \and
\authorblockN{Tuvi Etzion}
\authorblockA{Dept. of Computer Science\\
Technion-Israel Institute of Technology\\
Haifa 32000, Israel \\
Email: etzion@cs.technion.ac.il}}

\maketitle
\begin{abstract}
The rank modulation scheme has been proposed for efficient writing
and storing data in non-volatile memory storage. Error-correction
in the rank modulation scheme is done by considering permutation codes.
In this paper we consider codes in the set of all permutations on~$n$ elements,
$S_n$, using the Kendall's $\tau$-metric. We prove that there are no perfect
single-error-correcting codes in~$S_n$, where $n>4$ is a prime or $4\leq n\leq 10$.
We also prove that if such a code exists for $n$ which is not a prime then
the code should have some uniform structure. We define some
variations of the Kendall's $\tau$-metric
and consider the related codes and specifically we
prove the existence of a perfect single-error-correcting code in $S_5$.
Finally, we examine the existence problem
of diameter perfect codes in $S_n$ and obtain a new upper bound
on the size of a code in $S_n$ with even minimum Kendall's $\tau$-distance.
\end{abstract}


\section{Introduction}
Flash memory is a non-volatile technology that is both electrically
programmable and electrically erasable. It incorporates a set of cells
maintained at a set of levels of charge to encode information.
While raising the charge level of a cell is an easy operation,
reducing the charge level requires the erasure of the whole block to which the cell belongs.
For this reason charge is injected into the cell over several iterations.
Such programming is slow and can cause errors since cells may be injected with extra unwanted charge.
Other common errors in flash memory cells are due to charge leakage and reading disturbance that
may cause charge to move from one cell to its adjacent cells.
In order to overcome these problems, the novel framework of \emph{rank modulation codes} was introduced in  \cite{JMSB09}.
In this setup the information is carried by the relative ranking of the
cells’ charge levels and not by the absolute values of the charge levels.
This allows for more efficient programming of cells, and coding by the ranking of the cells' levels
is more robust to charge leakage than coding by their actual values.
In this model codes are subsets of $S_n$, the set of all permutations on $n$ elements,
where each permutation corresponds to a ranking of $n$ cells' levels.
Permutation codes were mainly studied in this context using two metrics, the infinity metric
and the Kendall's $\tau$-metric.

Codes in $S_n$ under the infinity metric were considered in \cite{KLT10,ShTs10,TaSc10,TaSc12}.
Anticodes in $S_n$ under the infinity metric were considered in \cite{Klo11,ShTs11,TaSc11}.

In this paper we consider codes using the Kendall's $\tau$-metric and some variation of the
Kendall's $\tau$-metric.
Under the Kendall's $\tau$-metric, codes in $S_n$
with minimum distance $d$ should correct up to
$\left\lfloor\frac{d-1}{2}\right\rfloor$ errors that are caused by charge leakage and read disturbance.
A~comprehensive work on error-correcting codes in $S_n$
using the Kendall's $\tau$-metric~\cite{KeGi90},
is given in \cite{JSB10}. In that paper~\cite{JSB10} there is also a construction of single-error-correcting codes
using codes in the Lee metric. This method was generalized in \cite{BM10} for the construction of
$t$-error-correcting codes that are of optimal size, up to a constant factor, where $t$ is fixed.
In \cite{ZJB12}, systematic-error-correcting codes were proposed. In particular,
they constructed a systematic single-error-correcting
code in~$S_n$ of size $(n-2)!$, which is of optimal size,
assuming that a perfect single-error-correcting code does not exist.
But, they only prove the nonexistence of perfect single-error-correcting codes for $n=4$.

The first part of this paper is devoted to perfect single-error-correcting codes in $S_n$,
using the Kendall's $\tau$-metric and
related structures. Perfect codes is one of the most fascinating topics in coding
theory. A perfect code in a given metric is a code in which the
set of spheres with a given radius~$R$ around its codewords forms a
partition of the space. These codes were mainly considered for the
Hamming scheme,
e.g.~\cite{EtVa94,Mol86,Phe83,Phe84a,Phe84}. They were
also considered for other schemes such as the Johnson scheme, e.g.~\cite{Etz96,EtSc,Roos}, the
Grassmann scheme \cite{Chi87,MaZh95}, and to a larger extent in the Lee and the
Manhattan metrics, e.g.~\cite{Etz11,GoWe70,Hor09,Post}.
Perfect codes were also considered on Cayley graphs~\cite{DeSe03} and for
distance-transitive graphs~\cite{Big73}.

The rest of this work is organized as follows. In
Section~\ref{sec:basic} we define the basic concepts for
the Kendall's $\tau$-metric and for perfect codes.
In Section~\ref{sec:nonexist} we prove the nonexistence of
a perfect single-error-correcting code in $S_n$, using the Kendall's $\tau$-metric,
where $n>4$ is a prime or $4\leq n\leq 10$.
We also show that perfect single-error-correcting
codes must have a uniform structure. If we slightly modify the Kendall's
$\tau$-distance to define a cyclic Kendall's $\tau$-distance,
then we have at least one perfect single-error-correcting code in $S_5$.
This code and more variations of the Kendall's $\tau$-metric
are discussed in Section~\ref{sec:cyclic}. In Section~\ref{sec:diameter} we examine
diameter perfect codes in $S_n$, using the Kendall's $\tau$-metric,
and improve some known upper bounds
on the size of a code in $S_n$ with even minimum Kendall's $\tau$-distance.
We conclude in Section~\ref{sec:conclusion} where we also present some questions for future research.


\section{Basic Concepts}
\label{sec:basic}

Let $S_n$ be the set of all permutations on the set of $n$ elements $[n]=\{1,2,\ldots,n\}$.
We denote a permutation $\sigma\in S_n$ by $\sigma=[\sigma(1),\sigma(2),\ldots,\sigma(n)]$.
For two permutations $\sigma , \pi \in S_n$, their multiplication $\pi\circ \sigma$ is defined as the
composition of $\sigma$ on $\pi$, namely, $\pi\circ\sigma(i)=\sigma(\pi(i))$, for all $1\leq i\leq n$.
Note that this operation on $S_n$ is not commutative.
Given a permutation $\sigma\in S_n$, an \emph{adjacent transposition}, $(i,i+1)$, is an exchange of two
adjacent elements $\sigma(i),\sigma(i+1)$ in $\sigma$, for some $1\leq i\leq n-1$.
The result is the permutation $\pi=[\sigma(1),\ldots,{\sigma(i-1)},{\sigma(i+1)},\sigma(i),{\sigma(i+2)},\ldots,\sigma(n)]$.
The permutation $\pi$ can also be written as ${\pi=(i,i+1)\circ \sigma}$, where $(i,i+1)$ is the cycle decomposition
of the permutation $[1,2,\ldots,i-1,i+1,i,i+2,\ldots,n]$.

For two permutations $\sigma,\pi \in S_n$, the Kendall's $\tau$-distance between $\sigma$ and
$\pi$, $d_K(\sigma,\pi)$, is defined as the minimum number of adjacent transpositions needed
to transform $\sigma$ into $\pi$ \cite{KeGi90}.
The following expression for $d_K(\sigma,\pi)$ is well known (e.g. \cite{JSB10}, \cite{Knu98}).

\begin{small}
\begin{equation*}
d_K(\sigma,\pi)= \\ |\{(i,j)~:~\sigma^{-1}(i)<\sigma^{-1}(j)\wedge\pi^{-1}(i)>\pi^{-1}(j)\}|.
\end{equation*}
\end{small}

Given a metric space, one can define codes. We say that $\cC\subset S_n$ is an $(n,M,d)$ code if
$|\cC|=M$ and $d_K(\sigma,\pi)\geq d$ for every two permutations $\sigma,\pi$ in $\cC$ ($d$ is called the
\emph{minimum distance} of the code $\cC$).

For a given space $\cV$ with a
distance measure $d( \cdot , \cdot )$, a subset $C$ of $\cV$ is a \emph{perfect
code} with \emph{radius}~$R$ if for every element $x \in \cV$
there exists a unique codeword $c \in C$ such that $d(x,c) \leq
R$. For a point $x \in \cV$, the \emph{sphere} of radius~$R$
centered at $x$, $S(x,R)$, is defined
by $S(x,R) \deff \{ y
\in \cV ~:~ d(x,y) \leq R \}$. In all the spaces and metrics
considered in this paper
the size of a sphere does not depend on the center of the
sphere. This is a consequence from the fact that right multiplication
of permutations is an isometric operation related to the distance. If
$C$ is a code with minimum distance $2R+1$ and $S$ is a sphere
with radius $R$ then it is readily verified that
\begin{theorem}
\label{thm:sphere_bound} For a code $C$ with minimum distance
$2R+1$ and a sphere $S$ with radius $R$ we have $| C | \cdot |S | \leq | \cV |$.
\end{theorem}

\vspace{0.1cm}

Theorem~\ref{thm:sphere_bound} known as the \emph{sphere packing
bound}. In a code $C$ which attains the sphere packing bound, i.e.
$| C | \cdot |S | = | \cV |$, the spheres with radius $R$ around
the codewords of $C$ form a partition of $\cV$. Hence, such a code
is a perfect code. A perfect code with radius $R$ is also called a
\emph{perfect $R$-error-correcting code}.

%
%
%
%
%
%
%


\section{Nonexistence of Some Perfect Codes}
\label{sec:nonexist}

In this section we prove that there are no perfect single-error-correcting codes in $S_n$,
where $n$ is a prime greater than $4$ or $4\leq n\leq 10$.
For each $i$, $1\leq i\leq n$, let $S_{n,i}$ be the subset of $S_n$ which consists of all the
permutations $\sigma\in S_n$ for which $\sigma(i)=1$,
i.e., \emph{one} is in the $i$th position in $\sigma$. Clearly we have that $|S_{n,i}|=(n-1)!$.

Assume that there exists a perfect single-error-correcting code $\cC\subset S_n$. For each $i$, $1\leq i\leq n$, let
$$
\cC_i=\cC\cap S_{n,i}~~~~\text{ and }~~~~ x_i=|\cC_i|.
$$

We say that a codeword $\sigma\in \cC$ \emph{covers} a permutation $\pi\in S_n$ if $d_K(\sigma,\pi)\leq 1$.
Since $\cC$ is a perfect single-error-correcting code, it follows that
every permutation in~$S_{n,1}$ must be at distance at most one from
exactly one codeword of $\cC$ and this codeword must belong either to $\cC_1$ or $\cC_2$.
Every codeword $\sigma\in \cC_1$ covers exactly~${n-1}$ permutations in $S_{n,1}$.
It covers itself and the $n-2$ permutations in~$S_{n,1}$
obtained from $\sigma$ by exactly one adjacent transposition $(i,i+1)$, $1 < i < n$.
Each codeword $\sigma\in\cC_2$ covers exactly one permutation $\pi \in S_{n,1}$, $\pi=(1,2)\circ\sigma$.
Therefore, we have the following equation

\begin{equation}
\label{eq:1r1}
(n-1)x_1+x_2=(n-1)!~.
\end{equation}

Similarly, by considering how the permutations of~$S_{n,n}$ are covered by
codewords of $\cC$, we have that

\begin{equation}
\label{eq:1rn}
x_{n-1}+(n-1)x_n=(n-1)!~.
\end{equation}

For each $i$, $2\leq i\leq n-1$, each permutation in~$S_{n,i}$ is covered by
exactly one codeword that belongs to
either $\cC_{i-1}$, $\cC_{i}$, or $\cC_{i+1}$. Each codeword
$\sigma\in \cC_i$ covers exactly~${n-2}$ permutations in $S_{n,i}$.
It covers itself and the $n-3$ permutations in $S_{n,i}$
obtained from $\sigma$ by exactly one adjacent transposition $(j,j+1)$, where $j<i-1$ or $j>i$.
Each codeword in $\cC_{i-1}\cup \cC_{i+1}$ covers exactly one permutation from $S_{n,i}$.
Therefore, for each $i$, ${2\leq i\leq n-1}$, we have the equation

\begin{equation}
\label{eq:1regular}
x_{i-1}+(n-2)x_i+x_{i+1}=(n-1)!~.
\end{equation}

Let $\mathbf{x}=(x_1,x_2,\ldots,x_n)$ and let $\mathbf{1}$ denote the all-ones vector.
Equations (\ref{eq:1r1}), (\ref{eq:1rn}), and (\ref{eq:1regular}) can be written in matrix
form as

\begin{equation}
\label{eq:linSystem}
A\mathbf{x}=(n-1)! \cdot \mathbf{1},
\end{equation}
where $A=(a_{i,j})$ is defined by
\begin{footnotesize}
\begin{equation*}
A= \left( \scalemath{0.85}{\begin{array}{ccccccccc}
n-1 & 1 & 0& 0 & \cdots & 0 & 0&\ldots & 0 \\
1& n-2 & 1 &0 & \cdots & 0 & 0& \ldots &0\\
0 & 1 & n-2 & 1 & \cdots & 0 & 0& \ldots &0 \\
\vdots& \vdots & \vdots & \vdots & \ddots & \vdots & \vdots &\vdots  &\vdots \\
\vdots& \vdots & \vdots & \vdots & \ddots & \vdots & \vdots &\vdots  &\vdots \\
0 &\ldots&0 & 0 & \cdots & 1 & n-2 & 1 &0\\
0 &\ldots&0& 0 & \cdots &0 & 1 & n-2 & 1 \\
0 &\ldots&0 & 0 & \cdots  & 0 & 0& 1 & n-1
\end{array}}\right).
\end{equation*}
\end{footnotesize}

Since the sum of every row in $A$ is equal to $n$ it follows that the linear equation system
(\ref{eq:linSystem}) has a solution ${\mathbf{y}=\frac{(n-1)!}{n} \cdot \mathbf{1}}$.
We will show that if $n > 3$ then $A$ is a nonsingular matrix
and hence $\mathbf{y}$ is the unique solution of (\ref{eq:linSystem}), i.e.,
$\mathbf{x}=\mathbf{y}$.
To this end, we need the following lemma, that can be easily verified, and is also
an immediate conclusion of the well known Gerschgorin circle theorem \cite{Ger31}.

\begin{lemma}\label{lem:Gresh}
Let $B=(b_{i,j})$ be an $n\times n$ matrix. If $|b_{i,i}|>\sum_{j\neq i}|b_{i,j}|$
for all $i$, $1\leq i\leq n$,
then $B$ is nonsingular.
\end{lemma}


For $n>4$, we have that for each $i$, $1\leq i\leq n$, $a_{i,i}\geq n-2>2\geq \sum_{j\neq i} a_{i,j}$.
By Lemma~\ref{lem:Gresh} it follows that $A$
is nonsingular. For $n=4$ it can be readily verified that the matrix $A$ is nonsingular.
Hence, for all $n\geq 4$, $x_{i}=\frac{(n-1)!}{n}$ for all $1\leq i\leq n$.
If $n=4$ or $n$ is a prime greater than $4$, then $x_i$ is not an integer and therefore,
a perfect single-error-correcting code does not exist for these parameters.

By using similar methods, we prove the nonexistence of perfect single-error-correcting codes in $S_n$
for $n\in\{6,8,9,10\}$. For each of these cases, we obtained a system of linear equations, which we solved by computer.
For $n>11$, the system of linear equations is very large and the computer failed to solve it.
We summarize our result in the following theorem.

\begin{theorem}
\label{thm:nonexist}
There is no perfect single-error-correcting code in $S_n$, where $n>4$ is a prime or $4\leq n\leq 10$.
\end{theorem}

By similar methods we can also prove the following property of perfect single-error-correcting codes.
\begin{theorem}
\label{thm:regularity}
Assume that there exists a perfect single-error-correcting code $\cC\subset S_n$,
where $n>11$. If $r<\frac{n}{4}$,
then for every sequence of $r$ distinct elements of $[n]$,
$i_1,i_2,\ldots,i_r$, and for every set of $r$ positions $1\leq j_1<j_2<\ldots<j_r\leq n$,
there are exactly $\frac{(n-r)!}{n}$ codewords $\sigma \in\cC$, such that
$\sigma(j_\ell)=i_\ell$, for each $\ell$, $1\leq \ell \leq r$.
\end{theorem}

Theorem \ref{thm:regularity} implies that perfect
single-error-correcting codes must have some kind of symmetry.
This might be useful to rule
out the existence of these codes for other parameters as well.


\section{The Cyclic Kendall's $\tau$-metric}
\label{sec:cyclic}
In this section we discuss a new metric which naturally
risen in the context of the Kendall's $\tau$-metric.

Given a permutation $\sigma\in S_n$, a \emph{c-adjacent transposition} is either an adjacent transposition
or the exchange of the elements $\sigma(1)$ and $\sigma(n)$.

For two permutations $\sigma,\pi \in S_n$, the \emph{cyclic Kendall's $\tau$-distance} between $\sigma$ and
$\pi$, $d_{\kappa}(\sigma,\pi)$, is defined as the minimum number of c-adjacent transpositions needed
to transform $\sigma$ into $\pi$.

For example, if $\sigma=[1,2,3,4]$ and $\pi=[4,3,2,1]$, then $d_{\kappa}(\sigma,\pi)=2$, since two
c-adjacent transpositions are enough to change the permutation from $\sigma$ to $\pi$: $[1,2,3,4]\to [4,2,3,1]\to
[4,3,2,1]$, and we need at least two c-adjacent transpositions for this purpose.
Clearly, $d_{\kappa}(\sigma,\rho)\leq d_K(\sigma,\rho)$ and therefore, if $\cC$ has
minimum cyclic Kendall's $\tau$-distance $d$ then $\cC$ also has minimum Kendall's $\tau$-distance at least $d$.

By Theorem~\ref{thm:nonexist} there is no perfect single-error-correcting code in $S_5$, using the Kendall's $\tau$-distance.
However, there exists a perfect single-error-correcting code in $S_5$, using the cyclic Kendall's $\tau$-distance.
For example, the following 20 codewords form such a code.

\begin{center}

$[0,1,2,3,4], ~[0,2,4,1,3], ~[0,3,1,4,2],~[0,4,3,2,1]$\\
$[1,2,3,4,0], ~ [2,4,1,3,0], ~ [3,1,4,2,0],~ [4,3,2,1,0]$\\
$[2,3,4,0,1], ~ [4,1,3,0,2], ~ [1,4,2,0,3], ~ [3,2,1,0,4]$\\
$[3,4,0,1,2], ~ [1,3,0,2,4], ~ [4,2,0,3,1],~ [2,1,0,4,3] $\\
$[4,0,1,2,3], ~ [3,0,2,4,1], ~ [2,0,3,1,4], ~[1,0,4,3,2]$.
\end{center}

Note that the permutations in each column are cyclic
shifts of the first permutation in the column.
Moreover, the permutations in the first row are of the form
$[0,\alpha,2\alpha,3\alpha,4\alpha]$, where $1\leq \alpha\leq 4$,
and multiplication is taken modulo 5.
A similar code of size $n\cdot(n-1)$ can be formed for each prime $n>5$.
These codes can be represented by another related distance measure.
We consider the following equivalence relation $E$ on~$S_n$.
For two permutations $\sigma = [ \sigma_1 , \sigma_2 ,~ \ldots ~,\sigma_n ]$
and $\pi = [ \pi_1 , \pi_2 ,~ \ldots ~, \pi_n ]$ we have that
$(\sigma,\pi) \in E$ if there exist an integer $i$, $1 \leq i \leq n$,
such that $\sigma = [\pi_i ,\pi_{i+1} , ~\ldots ~, \pi_n ,\pi_1 , ~ \ldots ~, \pi_{i-1} ]$.
Clearly, $E$ is an equivalence relation on $S_n$ with $(n-1)!$
equivalence classes, each one of size $n$. Let $S^c_n$ denote
the set of these $(n-1)!$ equivalence classes. Two elements
of $S^c_n$ are at distance one if there exist two representatives
of the two equivalence classes whose Kendall's $\tau$-distance
in one. If $\cC$ is a code with minimum Kendall's $\tau$-distance (of $S^c_n$) $d$
then there exists a code $\cC'$ in $S_n$, of size $n |\cC|$, with
minimum cyclic Kendall's $\tau$-distance $d$. Moreover, there exists
also a code of size $|\cC|$ in $S_{n-1}$ with minimum cyclic Kendall's $\tau$-distance $d$.

The computations of related distances, constructions of codes,
and the structure of
the two related graphs are intriguing research topics.
The results will be presented only in the full version of this work.

\section{Diameter Perfect Codes}
\label{sec:diameter}

In all the perfect codes the minimum distance of the code is an
odd integer. If the minimum distance of the code $\cC$ is an even
integer then $\cC$ cannot be a perfect code. The reason is
that for any two
codewords ${c_1 , c_2 \in \cC}$ such that $d(c_1,c_2)=2 \delta$, there
exists a word~$x$ such that $d(x , c_1)=\delta$ and
$d(x,c_2)=\delta$. For this case another concept is used, a
diameter perfect code, as was defined in~\cite{AAK01}. This
concept is based on the code-anticode bound presented by
Delsarte~\cite{Delsarte}. An \emph{anticode}~$\cA$ of \emph{diameter}~$D$
in a space~$\cV$ is a subset of words from~$\cV$ such that $d(x,y)
\leq D$ for all $x,y \in \cA$.

\begin{theorem}
\label{thm:abound} If a code $\cC$, in a space $\cV$ of a distance
regular graph, has minimum distance $d$ and in an anticode $\cA$
of the space $\cV$ the maximum distance is $d-1$ then $|\cC |
\cdot |\cA| \leq | \cV |$.
\end{theorem}

\vspace{0.2cm}

Theorem~\ref{thm:abound} which is proved in~\cite{Delsarte} is
a generalization of Theorem~\ref{thm:sphere_bound}
and it can be applied to
the Hamming scheme since the related graph is distance regular.
It cannot be applied to the Kendall's $\tau$-metric since the
related graph is not distance regular if $n > 3$. This can be easily verified
by considering the three permutations $\sigma=[1,2,3,4,5,\ldots,n]$, $\pi=[3,1,2,4,5,\ldots,n]$,
and $\rho=[2,1,4,3,5,\ldots,n]$ in~$S_n$. Clearly,
$d_K(\sigma,\pi)=d_K(\sigma,\rho)=2$ and
there exists exactly one permutation~$\alpha$ for which
$d_K (\sigma,\alpha)=1$ and $d_K (\alpha,\pi)=1$, while there exist exactly two permutations
$\alpha,\beta$ for which $d_K (\sigma,\alpha)=1$, $d_K (\alpha,\rho)=1$,
$d_K (\sigma,\beta)=1$, and $d_K (\beta,\rho)=1$.
Fortunately, an alternative proof which was given in~\cite{AAK01}
and was modified in~\cite{Etz11}
will work for the Kendall's $\tau$-metric.

\begin{theorem}
\label{thm:Ahlswede}
Let $\cC_{\cD}$ be a code in $S_n$
with Kendall's $\tau$-distances between codewords taken from a set
$\cD$. Let $\cA \subset S_n$ and let $\cC'_{\cD}$ be the largest
code in $\cA$ with Kendall's $\tau$-distances between codewords taken from the
set $\cD$. Then
$$
\frac{|\cC_{\cD}|}{n!} \leq \frac{|\cC'_{\cD}|}{|\cA|} ~.
$$
\end{theorem}

\begin{cor}
\label{cor:anti_Kendall} Theorem~\ref{thm:abound} holds for the Kendall's
$\tau$-metric, i.e. if a code $\cC \subset S_n$, has minimum Kendall's
$\tau$-distance $d$ and in an anticode $\cA \subset S_n$ the maximum
Kendall's $\tau$-distance is $d-1$ then $| \cC | \cdot |\cA| \leq n!$.
\end{cor}

If there exists a code $\cC\subset S_n$ with minimum Kendall's $\tau$-distance $d=D+1$,
and an anticode $\cA$ with diameter $D$ such that $|\cC|\cdot|\cA|=n!$, then
$\cC$ is called a \emph{diameter perfect} code with diameter $D$. In that case, $\cA$ must be an
anticode with maximum distance (diameter) $D$ of largest size, and $\cA$ is called an
\emph{optimal} anticode of diameter $D$. Thus, it is important to determine
the optimal anticodes in $S_n$ and their sizes. Using the size of such optimal anticodes we can
obtain by Corollary \ref{cor:anti_Kendall} an upper bound on the size of the related code in $S_n$.

One can verify that any permutation $\sigma = [ \sigma (1) , \sigma (2) , \ldots , \sigma (n)]$
in $S_n$ and its reverse $[ \sigma (n) , \ldots , \sigma (2) , \sigma (1)]$
form a diameter perfect code with diameter $\binom{n}{2} -1$. An optimal
anticode with diameter $\binom{n}{2} -1$ consists of $\frac{n!}{2}$
permutations, one permutation from each pair of permutations,
$\pi$ and its reverse in $S_n$.

An intriguing question is whether a sphere with radius~$R$ in $S_n$,
using the Kendall's $\tau$-metric, is an optimal anticode of diameter $2R$.
Such types of questions for other metrics were considered in \cite{AhBl08}.
For $n=4$, the sphere with radius $1$ has size $4$ and it is an optimal anticode of diameter $2$.
There exists an optimal anticode of diameter $2$ in $S_4$, which is not a sphere with radius $1$.
For example, the set $\cA=\{[1,2,3,4],[2,1,3,4],[1,2,4,3],[2,1,4,3]\}$ is an optimal
anticode of diameter $2$. A similar example exists for an optimal anticode of size
$9$ and diameter $4$ in $S_4$. However, for $n\geq 5$, we have the following theorem.

\begin{theorem}
Every optimal anticode with diameter $2$ in~$S_n$, $n\geq 5$, using the Kendall's $\tau$-distance,
is a sphere with radius $1$, whose size is $n$.
\end{theorem}

Let $\cS$ be a set of permutations in $S_n$ and let $\pi\in S_n$. We define
$\cS\circ\pi=\{\sigma\circ \pi~:~\sigma\in \cS\}$.

\begin{theorem}
\label{thm:diameterAnticode2}
Let $e=[1,2,\ldots,n]$ be the identity permutation of~$S_n$,
$n \geq 4$. And let $\cS(e,1)$ be the sphere
of radius 1 centered at $e$. Then the set
$$
\cA=\cS(e,1)\cup \cS(e,1)\circ (1,2)
$$
is an optimal anticode of diameter $3$, whose size is ${2(n-1)}$.
\end{theorem}
\begin{cor}
If $\cC\subset S_n$ is a code with minimum Kendall's $\tau$-distance $4$, then
$$
|\cC|\leq \frac{n!}{2(n-1)}.
$$
\end{cor}
We conjecture that the largest anticode with maximum
Kendall's $\tau$-distance $2R$ is a sphere with radius~$R$
if $2R < \binom{n}{2}$. We conjecture that the largest anticode with maximum
Kendall's $\tau$-distance $2R+1 < \binom{n}{2}$ is
$\cS(e,R)\cup \cS(e,R)\circ (1,2)$.
The size of this anticode will be discussed
in the full version of this work. It implies a new bound
on the size of a code in $S_n$ with minimum Kendall's $\tau$-distance $2R+2$.


\section{Conclusions and Open Problems}
\label{sec:conclusion}
We have considered several questions regarding perfect codes in
the Kendall's $\tau$-metric. We gave a novel technique to exclude
the existence of perfect codes using the Kendall's $\tau$-metric.
We applied this technique to prove that there are no perfect
single-error-correcting codes in $S_n$, where $n >4$ is a prime or
$4 \leq n \leq 10$, using the Kendall's $\tau$-metric.
We also proved that if such a code exists for other values of $n$ it
should have some uniform structure. We showed that if we use
a cyclic Kendall's $\tau$-metric then a perfect single-error-correcting
code exists in $S_5$. Finally, we examine the existence question
of diameter perfect codes in $S_n$. We obtained a new upper bound
on the size of a code in $S_n$ with even Kendall's $\tau$-distance.
Our discussion raises many open
problems from which we choose a few as follows.

\begin{enumerate}
\item Prove the nonexistence of perfect codes in $S_n$, using the
Kendall's $\tau$-metric, for more values of~$n$ and/or other distances.

\item Do there exist more perfect codes in $S_n$ using the cyclic
Kendall's $\tau$-metric?

\item Examine the cyclic Kendall's $\tau$-metric for its properties,
find upper bounds on the size of codes with this metric, and
construct codes with this metric. The same should be done if we consider
the set of equivalence classes $S^c_n$ of the relation $E$.

\item Is the sphere with radius $R$ in $S_n$ always optimal
as an anticode with diameter $2R < \binom{n}{2}$ in $S_n$? If yes, when there
are other optimal anticodes with the same parameters which are not
spheres?

\item What is the size of an optimal anticode in $S_n$ with
diameter $D$?

\item Improve the bounds on the size of codes in $S_n$
with even minimum Kendall's $\tau$-distance.
\end{enumerate}

\section*{Note added}
Theorem~\ref{thm:nonexist} is a special case of Theorem~5 in~\cite{DeSe03}
on perfect codes in Cayley graphs. But, the proof
of Theorem~5 in~\cite{DeSe03}
is wrong and some bounds on codes implied from it are false
with infinite counterexamples.

\section*{Acknowledgment}
This work was supported in part by the U.S.-Israel
Binational Science Foundation, Jerusalem, Israel, under
Grant No. 2012016. Sarit Buzaglo would like to thank Amir
Yehudayoff for many useful discussions. The authors also thank
one of the reviewers for bringing~\cite{DeSe03} to our attention.


\end{document}